\documentclass{elsart41}
\usepackage{graphics}
\usepackage{graphicx}
\usepackage{amssymb}
\usepackage{amsmath}

\begin{document}
\begin{frontmatter}



\title{A non-Hermitian analysis of strongly correlated quantum systems}
%

\author[AA]{Yuichi Nakamura\corauthref{Name1}},
\ead{yuichi@iis.u-tokyo.ac.jp}\
\author[BB]{Naomichi Hatano}

\address[AA]{Department of Physics, University of Tokyo, Komaba, Meguro, Tokyo, 153-8505, Japan}  
\address[BB]{Institute of Industrial Science, University of Tokyo, Komaba, Meguro, Tokyo, 153-8505, Japan}

\corauth[Name1]{Corresponding author. Tel: +81-3-5452-6156
fax: +81-3-5452-6155}

\begin{abstract}
We study a non-Hermitian generalization of strongly correlated quantum systems in which the transfer energy of electrons is asymmetric. It is known that a non-Hermitian critical point is equal to the inverse localization length of a Hermitian non-interacting random electron system. We here conjecture that we can obtain in the same way the correlation length of a Hermitian \textit{interacting non-random} system. We confirm the conjecture using exact solutions and numerical finite-size data of the Hubbard model and the antiferromagnetic $XXZ$ model in one dimension.
\end{abstract}

\end{frontmatter}

We study a non-Hermitian generalization of strongly correlated quantum systems by adding an imaginary vector potential $i\mathbf{g}$ (where $\mathbf{g}$ is a real constant) to the momentum operator. 
Hatano and Nelson~\cite{Hatano} analyzed this kind of non-Hermitian generalization of the one-electron Anderson model. Their purpose of the non-Hermitian generalization was to obtain a \textit{length} scale inherent in the wave function of the Hermitian model, namely the localization length, only from the non-Hermitian energy spectrum. We here conjecture that we can obtain in the same way the correlation length of a Hermitian \textit{interacting non-random} model.

The non-Hermitian one-electron Anderson model is given by~\cite{Hatano} 
\begin{align}
\mathcal{H}=&-t\sum_{x=1}^{L}\left(e^g| x+1\rangle\langle x|+e^{-g}| x\rangle\langle x+1|\right)\notag\\
&+\sum_{x=1}^{L}V_{x}|x\rangle\langle x|
\label{Anderson}
\end{align}
in one dimension, where $g$ is the parameter that generates the non-Hermiticity, corresponding to the imaginary vector potential, and $V_{x}$ is a random potential at site $x$.
At the Hermitian point $g=0$, all eigenvalues of the Hamiltonian~(\ref{Anderson}) are real. 
As we increase the non-Hermiticity $g$, a pair of eigenvalues collide at a point $g=g_{\mathrm{c}}$ and become complex. It was revealed~\cite{Hatano} that the non-Hermitian critical point $g_{\mathrm{c}}$ is equal to the inverse localization length of the eigenfunction of the original Hermitian Hamiltonian. 

The above study was for a non-interacting random electron system.
What length scale emerges in the energy spectrum if we consider the non-Hermitian generalization of an \textit{interacting non-random} system? We conjecture here that the answer is the correlation length. We show for the Hubbard model and the antiferromagnetic $XXZ$ model in one dimension that the non-Hermitian critical point of the ground state where the energy gap vanishes is equal to the inverse correlation length of the Hermitian point. We also analyze numerical data of non-Hermitian quantum systems of finite size, confirming the conjecture.

We first consider a one-dimensional non-Hermitian Hubbard model in the form
\begin{align}
\mathcal{H}=& -t\sum_{l,\sigma=\uparrow,\downarrow} (e^{g}c_{l+1,\sigma}^\dag c_{l,\sigma}+e^{-g}c_{l,\sigma}^\dag c_{l+1,\sigma})\notag\\
&+U\sum_{l} c_{l,\uparrow}^\dag c_{l,\uparrow}c_{l,\downarrow}^\dag c_{l,\downarrow}.\label{Hubbard_charge}
\end{align}
First, the inverse correlation length $1/\xi$ of the charge excitation at the Hermitian point $g=0$ was obtained by Stafford and Millis~\cite{Stafford} in the form
 \begin{equation}
\frac{1}{\xi}=\mathrm{arcsinh}(U/4t)-2\int_{0}^{\infty}\frac{J_{0}(\omega)\sinh((U/4t)\omega)}{\omega(1+e^{2(U/4t)\omega})}d\omega,\label{length}
\end{equation}
where $J_0(\omega)$ is the Bessel function of the first kind. 
Next, Fukui and Kawakami~\cite{Fukui} solved the non-Hermitian Hubbard model~(\ref{Hubbard_charge}) exactly. They showed that, as we increase the non-Hermiticity $g$, the Hubbard gap due to the charge excitation vanishes at a point $g_{\mathrm{c}}$ given in the form
\begin{equation}
g_{\mathrm{c}}=\mathrm{arcsinh}(U/4t)+2i\int_{-\infty}^{\infty}\arctan\frac{\lambda+iU/4t}{U/4t}\sigma(\lambda)d\lambda,\label{gc}
\end{equation}
where $\sigma(\lambda)$ is given by
\begin{equation}
\sigma(\lambda)=\frac{1}{2\pi}\int_{0}^{\infty}\mathrm{sech}\left(\frac{U}{4t}\omega \right)\cos(\lambda\omega)J_0(\omega)d\omega.
\end{equation}
We can show after some algebra that the expressions of the inverse correlation length $1/\xi$ in Eq.~(\ref{length}) and the non-Hermitian critical point $g_{\mathrm{c}}$ in Eq.~(\ref{gc}) are equal.%

We next consider the one-dimensional $S=1/2$ antiferromagnetic non-Hermitian $XXZ$ model in the form
\begin{equation}
\mathcal{H}=J\sum_{l}\left[\frac{1}{2}(e^{2g} S_{l}^{+}S_{l+1}^{-}+ e^{-2g} S_{l}^{-}S_{l+1}^{+})+\Delta S_{l}^{z}S_{l+1}^{z} \right]. \label{non-Herm XXZ}
\end{equation}
The non-Hermitian Hamiltonian~(\ref{non-Herm XXZ}) for $\Delta=1$ is derived by considering the second-order perturbation of the non-Hermitian Hubbard Hamiltonian 
\begin{align}
\mathcal{H}=&-t\sum_{l}(e^{g}c_{l+1,\uparrow}^\dag c_{l,\uparrow}+e^{-g}c_{l,\uparrow}^\dag c_{l+1,\uparrow}+e^{-g}c_{l+1,\downarrow}^\dag c_{l,\downarrow}\notag\\
&+e^{g}c_{l,\downarrow}^\dag c_{l+1,\downarrow})+U\sum_{l}c_{l,\uparrow}^\dag c_{l,\uparrow} c_{l,\downarrow}^\dag c_{l,\downarrow}\label{Hubbard_spin}
\end{align}
with respect to $t$.
As we increase the non-Hermiticity $g$ in Eq.~(\ref{Hubbard_spin}), we can eliminate the spin gap rather than the charge gap; see the difference between Eqs.~(\ref{Hubbard_charge}) and (\ref{Hubbard_spin}).
We then generalize the model to an arbitrary $\Delta$.  
Albertini \textit{et. al}~\cite{Albertini} calculated for $\Delta>1$ the non-Hermitian critical point $g_{\mathrm{c}}$ at which the spin gap vanishes:
\begin{equation}
g_{\mathrm{c}}=\frac{\gamma}{2}+\sum_{n=1}^{\infty}\frac{(-1)^{n}}{n}\tanh(n\gamma),\label{length_XXZ}
\end{equation}
where $\gamma=\text{arccosh}\Delta$. 
The expression (\ref{length_XXZ}) is actually well known~\cite{Baxter} as the inverse correlation length at the Hermitian point $g=0$. We have thus confirmed our conjecture for two solvable models. 

Unfortunately, it is difficult to know the ground-state properties of the non-Hermitian models (\ref{Hubbard_charge}) and (\ref{non-Herm XXZ}) for $g>g_{\mathrm{c}}$. However, we expect that the ground-state energy becomes complex in the region $g>g_{\mathrm{c}}$ on the basis of finite-size data, which we discuss below.

\begin{figure}
  \begin{center}
  \includegraphics[width=7.5cm,clip]{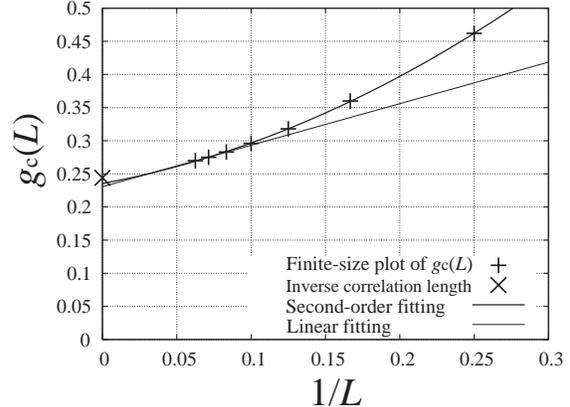} 
  \caption[]{The extrapolation of $g_{\mathrm{c}}(L)$ for the $XXZ$ model~(\ref{non-Herm XXZ}) with $\Delta=3$.
  The exact solution of the inverse correlation length is 0.244. The estimate $g_{\mathrm{c}}(\infty)$ by second-order fitting for $L=4,6,\dots,16$ is $0.235$ and that by linear fitting for $L=12,14,16$ is $0.231$.}
  \label{XXZ_scaling}
 \end{center}
 \end{figure}

We numerically diagonalized the Hamiltonians (\ref{Hubbard_charge}) and (\ref{non-Herm XXZ}) for finite size $L$. As we increase the non-Hermiticity $g$, the energy gap between the ground state and a low-lying excited state gradually decreases and eventually vanishes at $g=g_{\mathrm{c}}(L)$. Then the two eigenvalues become complex. This behavior is similar to that of the random Anderson model~\cite{Hatano}. We evaluated $g_{\mathrm{c}}(\infty)$ by extrapolating the finite-size data of  $g_{\mathrm{c}}(L)$. Figure~\ref{XXZ_scaling} shows, for example, extrapolation of $g_{\mathrm{c}}(L)$ for the $XXZ$ model~(\ref{non-Herm XXZ}) with $\Delta=3$. The estimate $g_{\mathrm{c}}(\infty)$ is indeed close to the inverse correlation length (\ref{length_XXZ}). It is noteworthy that we obtain a relatively good estimate of the inverse correlation length from the data of $g_{\mathrm{c}}(L)$ for small systems. 

We also carried out the above procedure of numerical analysis for frustrated quantum spin systems, including the Majumdar-Ghosh model~\cite{MG}. Details may be published elsewhere.
 
N. H. acknowledges financial support from the Sumitomo foundation.

\end{document}